\documentclass[aps,prl,amsmath,amssymb,superscriptaddress,preprintnumbers,twocolumn,10pt,nofootinbib,floatfix]{revtex4-2}
\pdfoutput=1

\usepackage[T1]{fontenc}
\usepackage[utf8]{inputenc}
\usepackage{graphicx}
\usepackage{bm}
\usepackage{booktabs}
\usepackage{amsmath}
\usepackage{amssymb}
\usepackage{xcolor}
\definecolor{royalblue}{rgb}{0.25,0.41,0.88}
\usepackage{hyperref}
\hypersetup{colorlinks=true,linkcolor=royalblue,citecolor=magenta,urlcolor=royalblue}

\newcommand{\dd}{\mathrm{d}}
\newcommand{\ii}{\mathrm{i}}
\newcommand{\Msun}{M_\odot}
\newcommand{\mfit}{\mathcal{R}}
\newcommand{\rmis}{\mathcal{M}}
\newcommand{\hsrc}{\tilde h_{\rm src}}
\newcommand{\psires}{\delta\psi_{\rm res}}
\newcommand{\tas}{\tau_{\rm as}}

\begin{document}

\title{Probing Dark Matter Substructure with Wave-Optics Distortions of Strongly Lensed LISA Gravitational Waves}

\author{Tonghua Liu}
\affiliation{School of Physics and Optoelectronic Engineering, Yangtze University, Jingzhou 434023, China}

\author{Kai Liao}
\email{liaokai@whu.edu.cn}
\affiliation{School of Physics and Technology, Wuhan University, Wuhan 430072, China}

\author{Marek Biesiada}
\affiliation{National Centre for Nuclear Research, Pasteura 7, PL-02-093 Warsaw, Poland}

\author{Jieci Wang}
\email{jcwang@hunnu.edu.cn}
\affiliation{Department of Physics, and Collaborative Innovation Center for Quantum Effects and Applications, Hunan Normal University, Changsha 410081, China;}
\affiliation{Hunan Research Center of the Basic Discipline for Quantum Effects and Quantum Technologies, Hunan Normal University, Changsha, Hunan 410081, China;}

\begin{abstract}
Strong lensing changes the phase of a gravitational-wave signal as well as its amplitude and arrival time.  We study whether this phase information can distinguish between three dark-matter structures in the lens: a Navarro--Frenk--White halo (NFW), a self-interacting dark matter (SIDM) halo, and a fuzzy-dark-matter field (FDM).  We generate waveforms for the detectable lensed massive-black-hole-binary population of a four-year LISA mission and fit every signal with the same smooth singular-isothermal-ellipsoid lens with external shear.  In 132 lens systems, 311 images are resolved as separate signals in time.  NFW and SIDM produce real waveform changes, but their slowly varying part is largely degenerate with the constant, gradient, and {Hessian of the smooth Fermat potential at the image.}
The coherent density fluctuations of FDM leave a larger frequency-dependent residual after this fit.  The NFW--FDM and SIDM--FDM populations become distinguishable with about 60 and 110 resolved image waveforms, respectively.  These results show that repeated lensed LISA signals can probe the spatial form of dark matter in lens galaxies, rather than only the total lensing mass.
\end{abstract}

\maketitle

\section{Introduction}

The 
{distribution} of dark matter is most uncertain on scales where galaxies contain little or no light.  Cold dark matter (CDM), warm dark matter (WDM), compact objects, self-interacting dark matter (SIDM), and ultralight scalar fields can agree on large-scale structure while predicting different mass distributions below the scale of luminous satellites \cite{2005PhR...405..279B,2017ARA&A..55..343B,2016PhR...643....1M,2018PhR...730....1T}.  Strong lensing is one of the few ways to probe dark matter on these scales.  In {strong lenses observed in the electromagnetic domain}, substructure is inferred from flux-ratio anomalies, image shifts, time delays, and perturbed arcs \cite{1998MNRAS.295..587M,2001ApJ...563....9M,2002ApJ...572...25D,2009MNRAS.392..945V,2014MNRAS.442.2017V,2024SSRv..220...58V,2019MNRAS.487.5721G,2020MNRAS.491.6077G,2020PhRvD.102f3502S,2024MNRAS.533.1687G}.
These measurements are useful, but they are affected by the source and by the lens galaxy.  Flux ratios can be changed by stars, dust, intrinsic source variability, and propagation through the lens.  Extended arcs contain more information, but require source reconstruction and a careful treatment of the lens light.  

The Laser Interferometer Space Antenna (LISA) provides a different measurement {of strongly lensed gravitational wave signal. Namely, }
the detector receives repeated copies of the same chirping signal, so the lens can be tested through the phase of the waveform rather than through the brightness of a resolved source.  Relative delays, magnifications, and Morse phases are encoded in the signals \cite{1999PThPS.133..137N,2003ApJ...595.1039T,2017arXiv170204724D,2018MNRAS.480.3842O,2019ApJ...874L...2H,2025NatAs...9..916S}.  Massive black-hole binaries remain in the LISA band for a long time, so a small additional lensing delay can produce a measurable phase change \cite{2017arXiv170200786A,2019CQGra..36j5011R,2016PhRvD..93b4003K,2016JCAP...04..002T,2021PhRvD.103h3011M}.  Previous wave-optics studies shown that LISA can be sensitive to low-mass halos and to the substructure in lens galaxies \cite{2023PhRvD.108l3543C,2023PhRvD.108j3529T,2026arXiv260304267A,2026arXiv260621519A}.

The relevant observable is the part of the change that a smooth lens cannot reproduce.  In the lensing potential, the constant term changes the arrival time, the first derivative shifts the image, and the second derivatives change the local convergence and shear.  These three effects are already 
{present in a} SIE-plus-shear lens, so they cannot by themselves identify the dark-matter model.  Across the small region around one image, an extended NFW halo is often close to this low-order form \cite{1997ApJ...490..493N,2015ApJ...799..108D}; its waveform effect is real, but much of it can be absorbed by refitting the smooth lens.  Gravothermal evolution can make an SIDM halo more centrally concentrated \cite{2000PhRvL..84.3760S,2018PhR...730....1T}.  FDM instead has density variations coherent over its de Broglie length \cite{2014NatPh..10..496S,2018MNRAS.478.2686C,2022MNRAS.517.1867L,2016PhR...643....1M,2017PhRvD..95d3541H,2025JCAP...07..025S,2024PhRvD.110h3536L}.  The {proper comparison should therefore be} between the part of each lensing potential that can be refitted and the part that changes across the image region in a way the smooth lens cannot follow.

Here we {perform such comparison on the simulated sample of detectable lensed massive-black-hole-binary population predicted for a four-year LISA mission in Ref.~\cite{2025PhRvD.112l3512G}.}
We draw {a sample of} singular-isothermal-ellipsoid (SIE) lenses with external shear, keep every detectable image whose signal is separated in time from its partners, and calculate the wave-optics response of NFW, SIDM, and FDM substructure \cite{2023PhRvD.108l3543C,2025PhRvD.111j3539V,2026arXiv260304267A,2026arXiv260621519A}.  The construction of the doubles, triples, and quadruples is given in Supplemental Material, Sec.~I.  
{One lensing system provides a one astrophysical event in which each resolved image supplies one waveform.}
After the delay and signal-duration cuts, the {final} sample contains 132 systems and 311 image waveforms.  We fit every image with the same smooth-lens waveform.  The 
{diagnostic} is the part of the dark-matter-induced waveform change that remains after the binary parameters have been refitted.
The precise NFW, SIDM, and FDM prescriptions, together with the SIDM profile
variation and the independent FDM density-realization check, are collected
in Supplemental Material, Sec.~IV.  Two checks test different physical
uncertainties: SIDM changes the inner halo profile at fixed mass, whereas FDM
changes the realization of a coherent density field at fixed boson mass.

\section{Wave-Optics Signal}

Wave optics becomes important when the gravitational-wave 
{period (inverse frequency)} is comparable to the lensing time-delay scale, so that the detector measures diffraction and interference rather than only geometric magnification and delay \cite{1999PThPS.133..137N,2003ApJ...595.1039T,2017arXiv170204724D,2023PhRvD.108l3543C,2025PhRvD.111j3539V}.  For macro image $i$, the smooth-lens waveform is
\begin{equation}
    \tilde h_i^{\rm macro}(f)=
    \sqrt{|\mu_i|}\exp(2\pi\ii f t_i-\ii\pi n_i)\hsrc(f),
    \label{eq:macro}
\end{equation}
where $f$ is the observed frequency, $\hsrc(f)$ is the unlensed source waveform, and $\mu_i$, $t_i$, and $n_i$ are the signed magnification, arrival time, and Morse index of image $i$.  We use $n_i=0,1/2,1$ for a minimum, saddle, or maximum image, respectively.  A small-scale perturbation changes the image waveform to
\begin{equation}
    \tilde h_i^{\rm DM}(f)=F_i(f)\tilde h_i^{\rm macro}(f),
    \label{eq:dmwave}
\end{equation}
with $F_i(f)$ the complex amplification factor produced by the small-scale structure.  A frequency-independent change can be absorbed into the image magnification and phase.  Substructure is identified by the frequency-dependent part of this change.
We write
\begin{equation}
F_i(f)=|F_i(f)|\exp[\ii\,\delta\phi_i(f)],
\end{equation}
so that $|F_i|$ is the frequency-dependent amplitude change and $\delta\phi_i$ is the frequency-dependent phase change.  The match in Eq.~(\ref{eq:match}) {below} is calculated from the full complex detector waveform and therefore includes both effects; the analysis is not phase-only.

Near a macro stationary point, let $\bm\eta=\bm\theta-\bm\theta_i$ be the angular displacement from the image position $\bm\theta_i$.  The smooth local time-delay surface is
\begin{equation}
T_{0,i}(\bm\eta)=\frac{1}{2}\bm\eta^{\rm T}{\bf H}_i\bm\eta ,
\end{equation}
where ${\bf H}_i$ is the Hessian matrix of the smooth Fermat potential at the image.  The perturbing potential is reduced to
\begin{equation}
\psires(\bm\eta)=
\psi_{\rm pert}(\bm\theta_i+\bm\eta)
-a_0-\bm a_1\!\cdot\!\bm\eta
-\frac{1}{2}\bm\eta^{\rm T}{\bf A}_2\bm\eta .
\label{eq:residualpsi}
\end{equation}
Here $a_0$ is the value of the perturbing potential at the image, $\bm a_1$ is its gradient, and ${\bf A}_2$ is its Hessian.  They describe, respectively, a change in arrival time, a shift of the image, and a change in local convergence and shear.  We remove these three terms before calculating the remaining wave effect.  They are the changes that the smooth lens is allowed to make; they are not extra dark-matter parameters in the fit.  The corresponding thin-lens expansion is given in Supplemental Material, Sec.~III.

For the time-separated images selected in our sample, the wave-optics modulation of one image is computed from the ratio
\begin{equation}
F_i(f)=
\frac{\int \dd^2\eta\,W(\eta)
e^{2\pi\ii f\tas[T_{0,i}(\eta)-\psires(\eta)]}}
{\int \dd^2\eta\,W(\eta)
e^{2\pi\ii f\tas T_{0,i}(\eta)}} .
\label{eq:localF}
\end{equation}
The window $W$ restricts the integral to the neighbourhood of image
$i$. The conversion factor $\tas$ is defined in
Eq.~(\ref{eq:tau_as}).  We use the same stationary point in the numerator and denominator.  A constant phase, an image displacement, and a quadratic change in the Fermat surface then cancel between the perturbed and smooth descriptions.  The remaining $F_i(f)$ is produced by structure that varies across the image neighbourhood.
The local expansion and the removal of the smooth-lens terms are derived in
Supplemental Material, Sec.~III; the model-specific tests of this procedure
are given in Sec.~IV.

{Wave effects become important when the dimensionless
frequency $w$ of the gravitational-wave signal is of order unity
\cite{2003ApJ...595.1039T,2019CQGra..36j5011R}.  For a lens at redshift
$z_L$, this parameter is}
\begin{equation}
w(f,M_L)
=
8\pi\frac{G M_L(1+z_L)f}{c^3},
\label{eq:wave_parameter}
\end{equation}
{where $G$ is Newton's constant, $c$ is the speed of
light, $M_L$ is the physical mass of the perturbing structure, and $f$
is the observed frequency.  Thus, the characteristic mass for
$w\simeq1$ is}
\begin{equation}
M_{\rm wave}\simeq
\frac{8.1\times10^6}{1+z_L}\,\Msun
\left(\frac{10^{-3}\,{\rm Hz}}{f}\right),
\label{eq:mw}
\end{equation}
{where $M_{\rm wave}$ serves only as a useful scale for
identifying the masses that can affect the LISA band.  For extended NFW, SIDM, and FDM structures,
$M_{\rm wave}$ is only a characteristic scale; the actual waveform
modulation depends on the projected mass distribution and the
frequency-dependent Fermat potential.} 
We therefore draw the CDM subhalo population over $10^5$--$10^7\Msun$, with $dN/dM\propto M^{-1.9}$ and projected substructure fraction $f_{\rm sub}=5\times10^{-4}$.  This is a conservative population of bound halos in the lens plane; halos along the line of sight would add further projected structure.  The chosen interval overlaps the mass range in which LISA wave-optics effects are expected to be important for highly magnified images \cite{2026arXiv260304267A}.  $M_{\rm wave}$ is the mass that is most effective at one frequency, not the mass of every halo in the population.  The NFW, SIDM, and FDM choices are defined in Supplemental Material, Sec.~IV.  Figure~\ref{fig:fdm} shows the FDM convergence fluctuation and the part of its Fermat phase that remains after the three smooth-lens terms $a_0$, $\bm a_1$ and ${\bf A}_2$  have been removed.

\begin{figure*}[t]
    \centering
    \includegraphics[width=\textwidth]{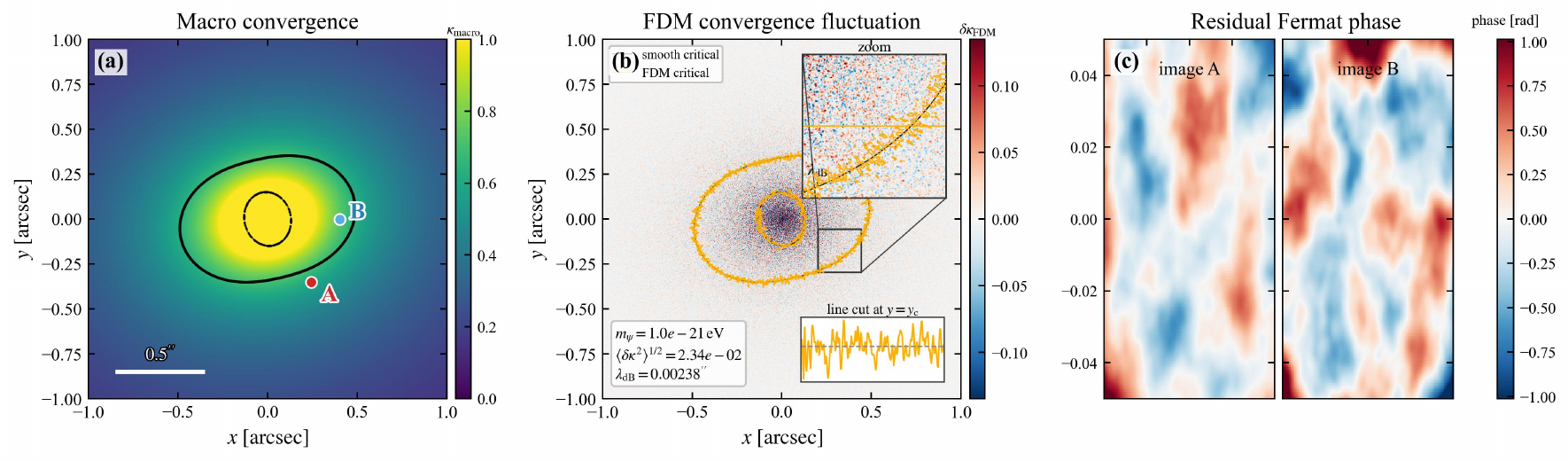}
    \caption{\textbf{FDM lensing geometry.}  (a) Convergence of the smooth SIE-plus-shear lens and the two macro images used in the calculation.  (b) Projected FDM convergence fluctuation with the smooth and perturbed critical curves.  (c) Residual Fermat phase in the neighbourhood of images A and B.}
    \label{fig:fdm}
\end{figure*}

\begin{figure}[t]
    \centering
    \includegraphics[width=\columnwidth]{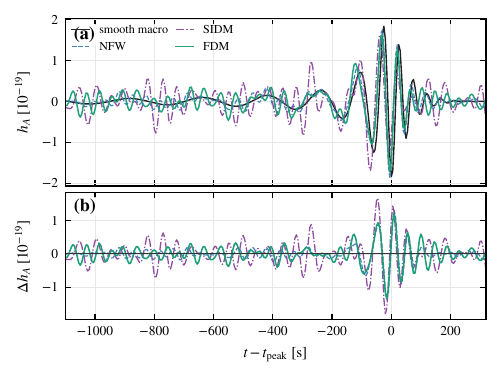}
    \caption{\textbf{Waveform change from dark-matter structure.}  The upper panel shows the TDI-$A$ strain near merger for the smooth macro image and for the same image with NFW, SIDM, or FDM structure.  The lower panel shows the difference from the smooth-lens waveform.}
    \label{fig:waveform}
\end{figure}

Figure~\ref{fig:fdm} shows how the FDM perturbation enters the calculation.  The SIE-plus-shear lens sets the large-scale critical curve and the positions of the macro images.  The FDM field is added on top of this lens, so the small corrugations in panel (b) are not an additional macro lens; they are the response of the same image to a spatially varying surface density.  Panel (c) shows the part of the Fermat phase that remains after the locally constant, linear, and quadratic pieces have been removed.  This distinction is important: the removed pieces change the overall delay, image position, convergence, or shear, whereas the remaining pattern changes across the integration region and produces a frequency-dependent factor $F_i(f)$.  The colour scale labelled rms is the root-mean-square amplitude of the plotted fluctuation.  The corresponding change in the measured waveform is shown in Fig.~\ref{fig:waveform}.

Figure~\ref{fig:waveform} gives a direct example of the effect 
{described} below.  The upper curves remain close because the dark-matter contribution is a perturbation to an already strongly magnified signal.  Their differences are clearer in the lower panel, where the residual changes sign and oscillates with time.  Such a residual is the time-domain form of a frequency-dependent change in $F_i(f)$; it cannot be described by changing only one magnification or one arrival time.  The Bayesian fit and the noise-weighted inner product determine how much of this difference remains after the binary parameters are allowed to adjust.

\section{Population Test}

We use the detectable lensed massive-black-hole-binary population of Ref.~\cite{2025PhRvD.112l3512G}, which contains 132 lensed systems over a four-year LISA mission.  For each source we draw an SIE lens with external shear \cite{1992grle.book.....S,1994A&A...284..285K,1997ApJ...482..604K}, solve for all macro images with \texttt{lenstronomy} \cite{2018PDU....22..189B,2021JOSS....6.3283B}, and keep every image that is detectable and separated in time from its partners.  Here $\Delta t$ is the delay between two macro images, and $T_{\rm sig}$ is the duration of the waveform segment used in the fit.  We require a detected partner with $1\,{\rm week}<\Delta t<1\,{\rm yr}$ and a separation larger than $\max(7\,{\rm days},T_{\rm sig})$.  The final sample has 311 image waveforms: 103 systems give two images, 11 give three, and 18 give four.  These images are separate signals in time, but images from the same lens share the same source and macro lens.  We therefore quote both numbers: 132 astrophysical systems and 311 waveform signals.  The median maximum image delay is 191 days and the median image signal-to-noise ratio, denoted by $\rho$, is 193.
The full image construction and the delay and signal-duration cuts are
given in Supplemental Material, Sec.~I; the corresponding system-level
resampling check is given in Sec.~VII.

\begin{figure*}[t]
    \centering
    \includegraphics[width=\textwidth]{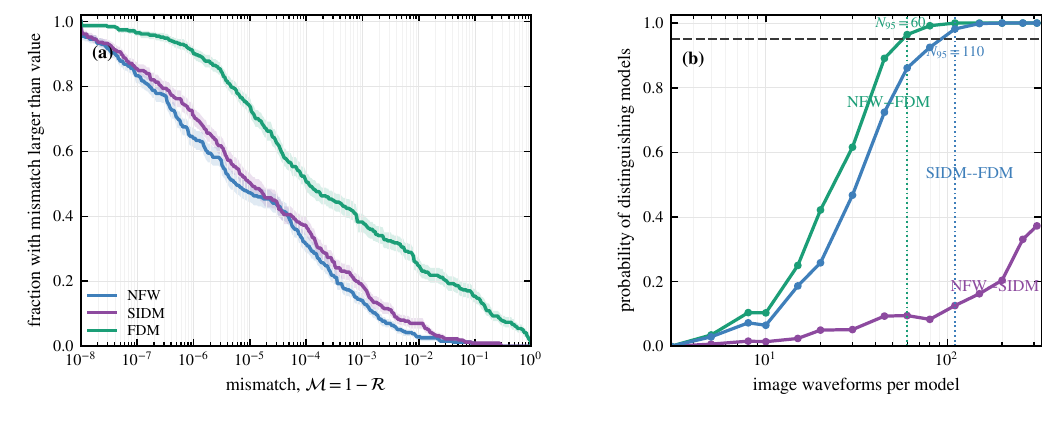}
    \caption{\textbf{Population mismatch and separation.}  (a) Cumulative fraction of image waveforms above a given mismatch $\rmis$.  The shaded bands show the variation under repeated catalogue draws.  (b) Fraction of repeated catalogue draws in which a two-sample Kolmogorov--Smirnov test separates two models, as a function of the number of resolved image waveforms per model.  The dotted lines mark the 95\% level.}
    \label{fig:match}
\end{figure*}

For each image we compute the detector waveform with an ESA trailing-orbit first-generation time-delay interferometry (TDI) response.  The source waveform is a frequency-domain massive-black-hole-binary waveform generated with \texttt{BBHx}, using the aligned-spin higher-harmonic template described in Supplemental Material, Sec.~V \cite{2018PhRvL.120p1102L,2020PhRvD.102b3033K}.  The injected signal, $h^{\rm DM}$, contains the dark-matter perturbation.  The fitted signal, $h^{\rm rec}$, uses the same binary waveform and the smooth SIE-plus-shear image factor, but has no small-scale structure.  The macro magnification, arrival time, and Morse phase come from the lens solution; their leading effects can also be adjusted by the effective distance, coalescence time, and coalescence phase in the waveform fit.  A superscript ``DM'' labels the injected signal and ``rec'' labels the reconstructed smooth-lens signal.  We apply the same fit to every selected image and calculate
\begin{equation}
\mfit=
\frac{(h^{\rm DM}|h^{\rm rec})}
{[(h^{\rm DM}|h^{\rm DM})(h^{\rm rec}|h^{\rm rec})]^{1/2}},
\label{eq:match}
\end{equation}
where the inner product is
\begin{equation}
(a|b)=4\,{\rm Re}\sum_{\alpha\in\{A,E,T\}}
\int_{f_{\rm min}}^{f_{\rm max}}
\frac{\tilde a_\alpha(f)\tilde b_\alpha^\ast(f)}{S_\alpha(f)}\,\dd f ,
\label{eq:inner}
\end{equation}
$\alpha$ labels the TDI channels, $\tilde a_\alpha(f)$ is the Fourier transform of the channel waveform, the star denotes complex conjugation, and $S_\alpha(f)$ is the one-sided LISA noise spectral density in that channel.  We use $f_{\rm min}=10^{-4}\,\mathrm{Hz}$ and $f_{\rm max}=5\times10^{-2}\,\mathrm{Hz}$ \cite{1998PhRvD..57.7089C,2003PhRvD..67b2001C,2017arXiv170200786A,2019CQGra..36j5011R}.  The construction of the $A$, $E$, and $T$ waveforms and their noise spectra is described in Supplemental Material, Sec.~VI.  A value of $\mfit\simeq1$ means that the smooth lens reproduces the perturbed waveform within the noise-weighted inner product.  We quote the corresponding mismatch,
\begin{equation}
    \rmis = 1-\mfit ,
    \label{eq:mismatch}
\end{equation}
which measures the residual disagreement after the smooth template has been allowed to adjust the binary parameters.  Thus $\mathcal{R}$ is the match and $\mathcal{M}=1-\mathcal{R}$ is the mismatch.  We evaluate $\mathcal{M}$ after fitting the eight binary parameters, rather than comparing two waveforms at fixed source parameters.  For each image, Fig.~\ref{fig:match} uses the mismatch at the posterior mode; the posterior samples provide the corresponding spread.  We use \texttt{bilby} with a Gaussian frequency-domain likelihood, first finding the best-fitting smooth waveform and then sampling the eight-parameter posterior with an affine-invariant Markov-chain sampler \cite{2019ApJS..241...27A,1997JGOpt..11..341S,2013PASP..125..306F}.  The waveform, frequency range, priors, and sampler settings are given in Supplemental Material, Sec.~V.
The separate smooth-lens control used to decide whether an individual residual is significant is given in Supplemental Material, Sec.~II.  It measures the mismatch produced by the same fitting procedure when no small-scale dark-matter perturbation is present.
The model-stability tests and the FDM mass dependence are described in
Supplemental Material, Secs.~IV and VIII, respectively.  The latter includes
both the full-catalogue mass scan and a representative conditional posterior
for $m_\psi$.

Figure~\ref{fig:match} contains the population result.  Panel (a) gives, at each value of $\mathcal M$, the fraction of image waveforms with a larger residual.  It is a population distribution, not an uncertainty bar for one event.  The median values for NFW, SIDM, and FDM are $5.69\times10^{-6}$, $1.03\times10^{-5}$, and $1.12\times10^{-4}$; their 95th percentiles are $4.50\times10^{-3}$, $1.46\times10^{-2}$, and 0.694.  These numbers show that NFW does change the waveform, even though its typical residual is small.  Across one image region, much of the NFW perturbation is close to a constant, a gradient, or a quadratic change in the Fermat surface, and the smooth fit can absorb those pieces.  The centrally concentrated SIDM population shifts the distribution upward modestly.  FDM produces the largest high-$\mathcal M$ population because its coherent density variations change across the image region and are less well represented by the smooth lens.

Panel (b) turns this distribution into an observing requirement.  The vertical axis is the fraction of repeated catalogue draws for which the two populations are distinguished by a two-sample Kolmogorov--Smirnov (KS) test with p-value $p<0.05$.  The dotted line is the 95\% criterion; it is a statement about repeated catalogues, not about the probability assigned to an individual event.  The NFW--FDM distributions reach this level with about 60 resolved image waveforms per model, while SIDM--FDM requires about 110.  If all images from one lens are retained together, the corresponding numbers are about 30 and 50 lens systems.  The distinction matters because images from one lens are separate signals in time but share the same source and macro lens.  NFW and SIDM remain closer to one another because they use the same abundance and mass function and differ mainly in the inner halo profile, whereas FDM changes the spatial pattern of the lensing potential itself.

\section{Discussion}

{Successful detections of strongly lensed gravitational signals by LISA will turn dark-matter substructure studies into a GW phase measurement.}
Figure~\ref{fig:waveform} shows that NFW, SIDM, and FDM all change the waveform.  Figure~\ref{fig:macroabs} shows why the NFW residual is often smaller after the constant, gradient, and curvature of the local lensing potential have been refitted.  FDM leaves more of its phase pattern because its density varies coherently across the image region.  The comparison therefore probes the spatial form of the lensing potential, not only the amount of dark matter.

{The simulated sample we used in our study contained both double and quadruple lenses.  We retained their images when the delays were long enough for the signals to be treated separately.  The result was 132 astrophysical lens systems and 311 time-resolved image waveforms.  We reported both counts because images from one system are separate measurements in time. However, they were not independent draws of a new lens. Our study was focused on a definite physical question: which part of a dark-matter-induced phase change cannot be explained by a smooth lens.  Under the wave-optics and population assumptions made in this paper, the answer differs between a conventional NFW-like halo and a coherent FDM field.}  A future mission analysis can extend this calculation by fitting all images from one lens together and by including line-of-sight halos, baryonic perturbers, alternative macro-lens models, waveform uncertainties, and the full LISA noise covariance \cite{2017PhRvD..95d4028B,2018ascl.soft05021L}.  These additions will change the forecast, but they do not change the physical distinction tested here.

\textit{Acknowledgments.---}
This work was supported by National Key R$\&$D Program of China (No. 2024YFC2207400); the National Natural Science Foundation of China under Grants No. 12675061; the Polish National Science Centre grant 2023/50/A/ST9/00579.

\bibliographystyle{apsrev4-2}
\bibliography{refs}

\appendix
\begin{center}
{\large\bf Supplemental Material}
\end{center}

\section{I. Population Construction}
\label{app:population}

The source layer follows the detectable lensed LISA massive-black-hole-binary population in Ref.~\cite{2025PhRvD.112l3512G}, which gives 132 lensed systems over the mission duration.  The macro-lens layer is regenerated with SIE+external-shear lenses.  Each system is solved for all macro images, including doubles and quads.  We then apply the image-level selection used in the main text:
\begin{enumerate}
\item the image signal-to-noise ratio must satisfy the detection criterion used in the source population;
\item the system must have a detected partner {separated in time by} $1\,{\rm week}<\Delta t<1\,{\rm yr}$;
\item accepted images in the same system must be separated in time by more than $\max(7\,{\rm days},T_{\rm sig})$, where $T_{\rm sig}$ is the duration of the waveform segment used in the fit.
\end{enumerate}
The final sample has 132 lens systems and 311 resolved image waveforms.  The distribution of usable images per system is 103 doubles, 11 triples, and 18 quadruples.  The distinction between systems and image waveforms is important: the former is the astrophysical event rate, while the latter is the number of time-resolved signals entering the waveform statistic.  Images from one lens share the same source and macro potential; the image-level result in Fig.~\ref{fig:match} treats them as separate waveform measurements, while the system-level calculation keeps their common origin.

\section{II. Smooth-Lens Null Range}
\label{app:null}

As a control, we also inject waveforms lensed only by the smooth SIE-plus-shear model and fit them in exactly the same way.  This is the null case: the signal is lensed, but it contains no small-scale dark-matter perturbation.  Its match is not exactly one because the signal has finite signal-to-noise ratio, the binary parameters are correlated, and the fit is numerical.  The resulting SNR-dependent range of $\mfit$ and $\rmis$ sets the range of mismatch expected from the smooth lens and the fitting procedure alone.  A dark-matter perturbation is relevant in this test only when its mismatch exceeds this reference range.

\begin{figure*}[t]
    \centering
    \includegraphics[width=\textwidth]{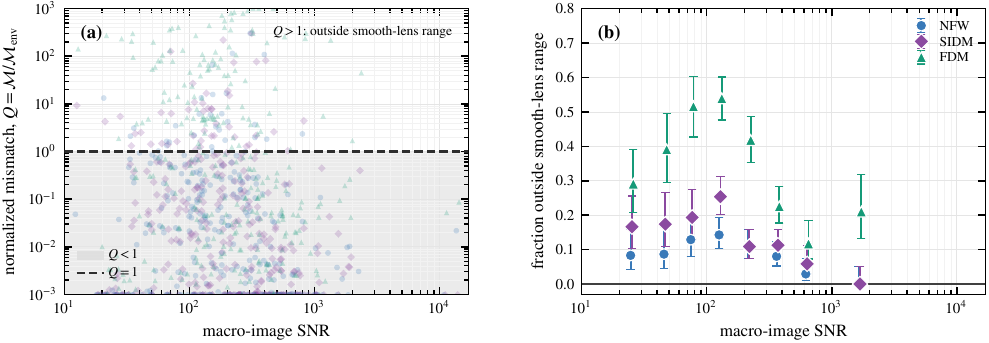}
    \caption{\textbf{Images outside the smooth-lens reference range.}  (a) Individual image waveforms plotted using the diagnostic ratio $Q=\mathcal{M}/\mathcal{M}_{\rm env}$.  The dashed line marks $Q=1$.  (b) Fraction of images above this boundary in logarithmic SNR bins; vertical bars are 68\% binomial intervals.}
    \label{fig:snrmismatch}
\end{figure*}

Figure~\ref{fig:snrmismatch} compares the selected images with this null range.
At fixed SNR, the mismatch varies because different images sample different
parts of the perturbing potential and because the smooth lens absorbs different
amounts of the perturbation.  We therefore define a comparison ratio
\begin{equation}
Q \equiv \frac{\mathcal{M}}{\mathcal{M}_{\rm env}(\rho)},
\end{equation}
where $\rho$ is the SNR of the image and $\mathcal{M}_{\rm env}(\rho)$ is the upper
edge of the smooth-lens reference range at that SNR.  Thus $Q<1$ means that
the residual is consistent with the smooth-lens null range, whereas $Q>1$
means that the signal contains more mismatch than the smooth macro lens can
account for.  $Q$ is used only to compare an image with the smooth-lens
control at the same SNR; {the mismatch
$\mathcal{M}$ shown in Fig.~\ref{fig:match} remains the proper population diagnostic}.

The two panels show the same physical test in complementary forms.  Panel (a)
retains every fitted image waveform.  The NFW points lie predominantly below
$Q=1$, whereas FDM produces a visibly larger group above the boundary.  Panel
(b) summarizes the same selection in logarithmic SNR bins; the binomial
intervals show the uncertainty in each catalogue fraction.  The FDM fraction
above the smooth-lens range is systematically higher than the NFW fraction over
the well-populated part of the sample, with SIDM between them.  These are
fractions of the simulated image catalogue, not posterior probabilities for an
individual event.  The result gives a direct physical check of the population
comparison: FDM more often leaves a phase-dependent residual that cannot be
absorbed by the smooth macro lens, whereas most NFW events remain compatible
with that smooth-lens description.

\section{III. Smooth-Lens Modes}
\label{app:smoothmodes}

The dimensional thin-lens time delay is
\begin{equation}
t_d(\bm\theta,\bm\beta)=
\frac{1+z_L}{c}\frac{D_LD_S}{D_{LS}}
\left[\frac{1}{2}|\bm\theta-\bm\beta|^2-\psi(\bm\theta)\right],
\end{equation}
where $z_L$ is the lens redshift, $D_L$, $D_S$, and $D_{LS}$ are angular-diameter distances to the lens, to the source, and from lens to source, $\bm\beta$ is the source position, and $\psi$ is the two-dimensional lensing potential \cite{1992grle.book.....S}.  Angular quantities in the analytic lensing equation are measured in
radians. In the numerical calculation, the local image-plane
coordinates are expressed in arcseconds. We therefore define
\begin{equation}
\tas =
\frac{1+z_L}{c}
\frac{D_LD_S}{D_{LS}}
\left(\frac{\pi}{648000}\right)^2 ,
\label{eq:tau_as}
\end{equation}
which converts a Fermat-potential difference expressed in
arcsec$^2$ into a physical time delay. This factor appears explicitly
in Eq.~(\ref{eq:localF}), so that
$f\tas[T_{0,i}(\bm\eta)-\psires(\bm\eta)]$ is dimensionless.  The potential in Eq.~(\ref{eq:residualpsi}) removes the three changes that a smooth lens can fit.  The scalar $a_0$ changes the absolute phase and arrival time, the vector $\bm a_1$ shifts the image, and the matrix ${\bf A}_2$ changes the local convergence and shear.  The remaining $\psires$ is not the full dark-matter potential; it is the part that varies beyond this quadratic form.

\begin{figure*}[t]
    \centering
    \includegraphics[width=\textwidth]{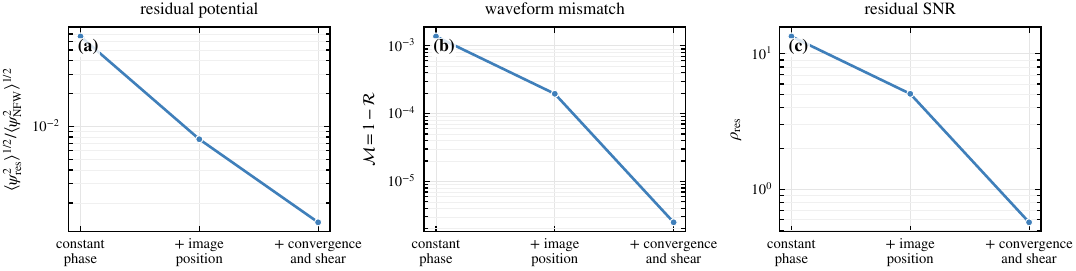}
    \caption{\textbf{Local smooth-lens terms.}  The same NFW-perturbed image is shown after successively removing a constant, a gradient, and a Hessian from the perturbing Fermat potential.  The panels show the residual potential, waveform mismatch, and residual signal-to-noise ratio.}
    \label{fig:macroabs}
\end{figure*}

Figure~\ref{fig:macroabs} isolates the origin of the small NFW residual.  An NFW subhalo changes the waveform, but its contribution varies slowly across one image region.  A constant {$a_0$} changes the absolute delay, a gradient changes the image position, and a Hessian changes the local convergence and shear.  These changes are already available to the smooth SIE-plus-shear lens, so the fit can reproduce much of the NFW contribution.  What remains in the last panel is the part that changes beyond this quadratic approximation.  Thus a small NFW mismatch means that the perturbation is largely degenerate with the local macro lens; it does not mean that NFW has no effect on the waveform.

\section{IV. Dark-Matter Models}
\label{app:models}

The NFW density profile is
\begin{equation}
\rho_{\rm NFW}(r)=
\frac{\rho_s}{(r/r_s)(1+r/r_s)^2}.
\end{equation}
Here $r$ is the three-dimensional distance from the halo centre, $r_s$ is the scale radius, and $\rho_s$ is the characteristic density.  {One often introduces the concentration parameter defined as $c_{\rm NFW}=r_{200}/r_s$, where $r_{200}$ denotes the radius of a sphere enclosing 200 times the critical density.}  The subhalo population uses $dN/dM\propto M^{-1.9}$ between $10^5$ and $10^7\Msun$, a concentration $c_{\rm NFW}$ scatter of 0.12 dex, and a projected substructure fraction $f_{\rm sub}=5\times10^{-4}$.  Here $N$ is the number of subhalos, $M$ is the subhalo mass, and $f_{\rm sub}$ is the fraction of the lens-plane surface density assigned to these halos.  This value is lower than the few-percent fractions sometimes inferred when more massive subhalos and line-of-sight halos are included, but is appropriate for a conservative bound-subhalo lens-plane population in the $10^5$--$10^7\Msun$ wave-optics interval \cite{2002ApJ...572...25D,2019MNRAS.487.5721G,2020MNRAS.491.6077G,2020PhRvD.102f3502S,2024MNRAS.533.1687G}.

The SIDM model uses the same halo masses, mass function, and abundance as NFW.  Self-interactions can make a central core, while gravothermal evolution in a dense or old halo can increase its central density \cite{2000PhRvL..84.3760S,2018PhR...730....1T,2022JCAP...09..054Y,2023ApJ...958L..39N,2025arXiv250214964H}.  We use this centrally concentrated branch.  To change only the inner profile, we keep the halo positions and total masses fixed and multiply the NFW concentration by
\begin{equation}
c_{\rm SIDM}=B_c c_{\rm NFW},\qquad B_c=3 .
\end{equation}
The reference value $B_c=3$ describes a moderately evolved, centrally concentrated SIDM halo.  It is not a measurement of a particular self-interaction cross section and does not describe every SIDM halo.  Since the halo masses and abundance are unchanged, the difference from NFW comes from the inner density profile, not from adding more lensing mass.  The Supplementary Material repeats the calculation for other values of $B_c$; a core-forming SIDM halo would be a different model.

\begin{figure}[t]
    \centering
    \includegraphics[width=\columnwidth]{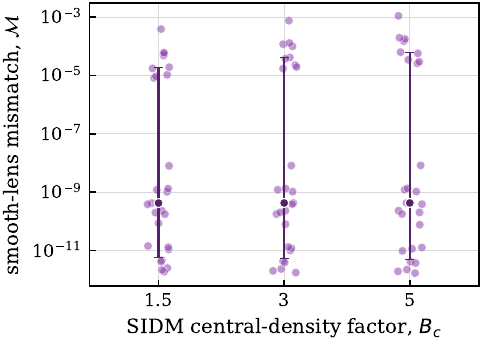}
    \caption{\textbf{SIDM profile variation.}  Points show individual image waveforms for three values of the central-density factor $B_c$; dark markers and vertical bars show the median and central 68\% interval.  The halo mass function, projected abundance, macro lens, and smooth-lens fit are unchanged.}
    \label{fig:sidmstability}
\end{figure}

\paragraph{Dependence on the SIDM central density.}
The reference SIDM choice is $B_c=3$.  We repeat the selected-image calculation at $B_c=1.5$, $3$, and $5$, keeping the halo masses, abundance, positions, macro lens, source waveform, detector response, and smooth-lens fit fixed.  Each value uses 27 image waveforms.  The median $\rmis$ values are $4.25\times10^{-10}$, $4.27\times10^{-10}$, and $4.29\times10^{-10}$; the corresponding 95th percentiles are $6.0\times10^{-5}$, $1.28\times10^{-4}$, and $1.94\times10^{-4}$.  Figure~\ref{fig:sidmstability} shows how changing the inner density changes the upper part of the mismatch distribution while leaving the typical event nearly unchanged.  The range of $B_c$ changes the strength of the SIDM residual without changing the halo abundance or total mass, and the result remains below the FDM population for the reference choice.

For FDM, the {relevant scale is set by the } de Broglie wavelength
\begin{equation}
\lambda_{\rm dB}\simeq
1.2\,{\rm kpc}
\left(\frac{10^{-22}{\rm eV}}{m_\psi}\right)
\left(\frac{100\,{\rm km\,s^{-1}}}{v}\right),
\end{equation}
where $m_\psi$ is the boson mass and $v$ is the characteristic velocity in the host halo.  We use $m_\psi=10^{-21}\,{\rm eV}$ for the main comparison.  For the adopted host normalization, the corresponding coherence scale is comparable to the region around a macro image sampled by the lensed LISA waveform \cite{2014NatPh..10..496S,2016PhR...643....1M,2017PhRvD..95d3541H,2025JCAP...07..025S}.  The dependence on this mass is tested over the full image catalogue in Supplemental Material, Sec.~VIII.  We describe the projected density fluctuation by the dimensionless convergence field $\delta\kappa_{\rm FDM}$ and obtain its lensing potential from
\begin{equation}
\nabla_\theta^2\delta\psi_{\rm FDM}=2\delta\kappa_{\rm FDM}.
\end{equation}
where $\nabla_\theta^2$ is the two-dimensional angular Laplacian and $\delta\psi_{\rm FDM}$ is the corresponding perturbing lensing potential.  The three models use the same macro-image sample and the same smooth-lens subtraction.  The FDM field amplitude is fixed by the host normalization throughout the population calculation.

\begin{figure}[t]
    \centering
    \includegraphics[width=\columnwidth]{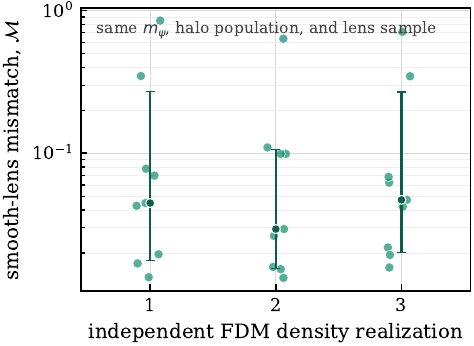}
    \caption{\textbf{FDM density-realization check.}  Individual image
    mismatches for three independent FDM density realizations at fixed
    $m_\psi$, halo population, macro lens, and detector response.  Dark
    points and vertical bars show the median and central 68\% interval for
    each realization.  This is a same-setup check on a representative
    nine-image subset; it does not replace the full-catalogue result.}
    \label{fig:fdmrealization}
\end{figure}

\paragraph{Dependence on the FDM density realization.}
The FDM density field is stochastic, so the population result should not
depend on one particular draw of that field.  As a separate stability check,
we generated three independent density realizations at the reference mass
$m_\psi=10^{-21}\,\mathrm{eV}$ and evaluated the same nine representative
image waveforms for each realization.  The halo population, projected
abundance, macro lens, source waveform, detector response, and smooth-lens
fit were unchanged.  The realization medians {of the mismatch $\rmis$ } are
$4.48\times10^{-2}$, $2.94\times10^{-2}$, and $4.71\times10^{-2}$, while the
combined 16--84\% range is $1.61\times10^{-2}$--$3.09\times10^{-1}$.
Figure~\ref{fig:fdmrealization} shows the individual image results and their
central 68\% ranges.  The spread is the expected variation with the sampled
line of sight and image position; the FDM residual remains substantially
    larger than the NFW-like residual in this comparison.  This test is
not used to change the event count in Fig.~\ref{fig:match}; it verifies that
the qualitative ordering is not tied to one random density realization.

\section{V. Bayesian Fit with a Smooth Lens}
\label{app:postmatch}

For each dark-matter model $q\in\{{\rm NFW,SIDM,FDM}\}$, we use a zero-noise injection, $d_q=h_q$, where $h_q$ is the detector response to the waveform perturbed by model $q$.  The data label $d_q$ is therefore a known signal in this calculation, not a random noise realization.  The Gaussian likelihood weights the difference between this signal and the fitted smooth-lens waveform with the LISA noise spectrum.  The fitted waveform is $h_{\rm rec}(\vartheta)$, with $\vartheta$ {being} the eight binary parameters listed below.  The source waveform is generated with \texttt{BBHx} as a frequency-domain massive-black-hole-binary waveform using an aligned-spin higher-harmonic template \cite{2018PhRvL.120p1102L,2020PhRvD.102b3033K}.  We include the $(2,2)$, $(2,1)$, $(3,3)$, $(3,2)$, $(4,4)$, and $(4,3)$ modes.  The sky position and polarization are fixed to the injected values, while the binary parameters are fitted.  The macro magnification, arrival time, and Morse phase come from the regenerated SIE-plus-shear image solution.  The constant, gradient, and quadratic parts of the perturbing Fermat potential are removed when calculating the wave-optics factor, so the fit tests the remaining frequency-dependent change without adding separate dark-matter parameters.

The sampler uses the eight variables
\begin{equation}
\vartheta_{\rm samp}=\{\ln m_{1,z},\ln m_{2,z},\chi_{1z},\chi_{2z},
\cos\iota,\ln D_L^{\rm eff},\phi_c,t_c\},
\label{eq:params}
\end{equation}
Here $m_{1,z}\ge m_{2,z}$ are detector-frame component masses, $\chi_{1z}$ and $\chi_{2z}$ are the dimensionless spin components parallel to the orbital angular momentum, $\iota$ is the inclination, $D_L^{\rm eff}$ is the effective luminosity distance after macro magnification, $\phi_c$ is the coalescence phase, and $t_c$ is the coalescence time.  For the corner plot we display the derived variables
\begin{equation}
\vartheta_{\rm plot}=\{\mathcal{M}_{c,z},q,\chi_{\rm eff},\chi_a,
\cos\iota,D_L^{\rm eff},\phi_c,t_c-\bar t_c\},
\label{eq:plotparams}
\end{equation}
where $\mathcal{M}_{c,z}=(m_{1,z}m_{2,z})^{3/5}/(m_{1,z}+m_{2,z})^{1/5}$ is the detector-frame chirp mass, $q=m_{2,z}/m_{1,z}$ is the mass ratio, $\chi_{\rm eff}=(m_{1,z}\chi_{1z}+m_{2,z}\chi_{2z})/(m_{1,z}+m_{2,z})$ is the mass-weighted aligned spin, $\chi_a=(\chi_{1z}-\chi_{2z})/2$ is the antisymmetric spin combination, and $\bar t_c$ is the median coalescence time of the posterior samples.  The plotted distance is in Gpc and the plotted chirp mass is in units of $10^6\Msun$.  The smooth image magnification, Morse phase, and arrival time are included through the macro-image factor in Eq.~(\ref{eq:macro}); the local quadratic terms have already been removed from the perturbing wave-optics calculation described in Sec.~III.  The likelihood is the standard Gaussian frequency-domain likelihood used in compact-binary parameter estimation,
\begin{equation}
{\cal L}\propto
\exp\left[-\frac{1}{2}
\left(d_q-h_{\rm rec}(\vartheta)\middle|
d_q-h_{\rm rec}(\vartheta)\right)\right],
\end{equation}
The inner product is evaluated from $10^{-4}$ to $5\times10^{-2}\,{\rm Hz}$.  We construct a $1.2\,{\rm yr}$ segment with cadence $\Delta t=10\,{\rm s}$ and use the same frequency sampling for the injection and the fit.  The same LISA noise curve, orbit, and TDI response are used for both.

\begin{figure*}[t]
    \centering
    \includegraphics[width=\textwidth]{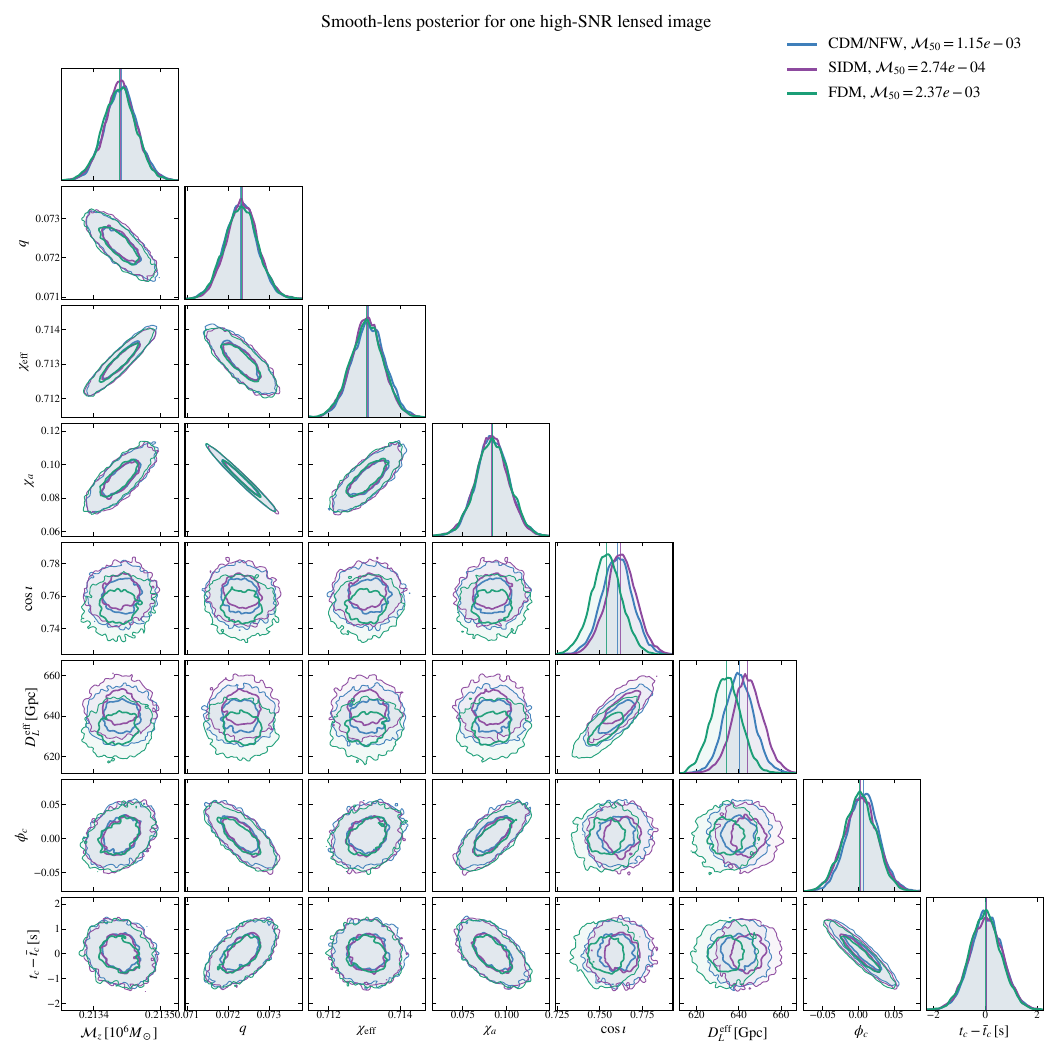}
    \caption{\textbf{Posterior from a smooth-lens fit.}  Posterior samples for one high signal-to-noise image injected with NFW, SIDM, or FDM structure and fitted with the same smooth SIE-plus-shear waveform.  The plot shows the derived variables in Eq.~(\ref{eq:plotparams}); colours identify the injected model.}
    \label{fig:posterior}
\end{figure*}

The priors are uniform and centred on the injected parameters. To make
the parameter-estimation procedure reproducible, we specify the prior
widths directly. The detector-frame component masses are allowed to vary
by $\pm4\%$ around their injected values, the aligned-spin components
have widths of $0.12$, and $\cos\iota$ has a width of $0.07$. The
effective luminosity distance is varied within a width of $0.05$ in
$\ln D_L^{\rm eff}$. The coalescence phase is given the range
$[-\pi,\pi]$, and the coalescence time has a width of
$0.004\,\mathrm{yr}$. All priors are truncated to the physical domains
of the corresponding parameters.  We use \texttt{bilby}.  We first find the best-fitting smooth waveform with differential evolution and a Nelder--Mead maximization \cite{2019ApJS..241...27A,1997JGOpt..11..341S}, then sample the eight-parameter posterior with 32 walkers for 240 steps, discard the first 80, and retain every fifth step \cite{2013PASP..125..306F}.  The same procedure is used for every NFW, SIDM, and FDM image.  Figure~\ref{fig:match} uses the mismatch at the posterior mode, namely the highest-density point of the fitted posterior; Fig.~\ref{fig:posterior} shows the parameter correlations for one representative image.

The smooth-lens posterior is
\begin{equation}
p(\vartheta|d_q,{\rm smooth})
\propto
{\cal L}\!\left[d_q|h_{\rm rec}(\vartheta)\right]
\pi(\vartheta),
\end{equation}
where $\pi(\vartheta)$ is the prior.  For each posterior sample we evaluate the same match defined in Eq.~(\ref{eq:match}), replacing $h^{\rm rec}$ by $h_{\rm rec}(\vartheta)$.  The value plotted in Fig.~\ref{fig:match} is the mismatch at the posterior mode; the posterior samples quantify the uncertainty around that value and are illustrated in Fig.~\ref{fig:posterior}.  Throughout the paper, a smaller $\mfit$ and a larger $\rmis$ mean that the smooth lens leaves a larger unmodeled waveform distortion.

Figure~\ref{fig:posterior} shows what the fit is allowed to change before $\mathcal M$ is evaluated.  The three posteriors are obtained from the same image, with the same detector response and the same eight-parameter waveform model; only the injected dark-matter structure is changed.  The shifts and correlations in masses, spins, distance, phase, and coalescence time show how the smooth template attempts to reproduce the lensing phase.  The mismatch that remains after this adjustment is the quantity used in Fig.~\ref{fig:match}.  The posterior plot is therefore a check of the fitting step, while the population comparison comes from the 311 image waveforms.

\section{VI. Detector Response}
\label{app:response}

The detector response uses the ESA trailing-orbit geometry in \texttt{lisatools} and first-generation TDI.  For each image, the source waveform is multiplied by the macro-image factor and by $F_i(f)$, then projected into the LISA $A$, $E$, and $T$ channels used by \texttt{BBHx}.  The likelihood uses the corresponding noise spectra $S_A(f)$, $S_E(f)$, and $S_T(f)$.  The same orbit, sky position, polarization, response, and noise curves are used for the injection and the smooth-lens fit.  The fit therefore compares the signals in the detector channels, not only in the source frame.  The purpose of this step is physical: a residual is counted only if it survives the mapping from the lensed source waveform to the LISA observables.

For the 311 image waveforms, the mean mismatches are $2.72\times10^{-3}$ for NFW, $4.81\times10^{-3}$ for SIDM, and $7.67\times10^{-2}$ for FDM.  The 90th percentiles are $1.71\times10^{-3}$, $3.59\times10^{-3}$, and 0.196.  The median, upper part of the distribution, and model-separation test all give the same ordering; this is the basis of Fig.~\ref{fig:match}(b).

\section{VII. From Image Signals to Complete Lens Systems}
\label{app:systemresampling}

Each accepted image is a separate waveform because it arrives at a different time, but images {coming form the same lensing system} 
are not independent astrophysical events.  We therefore repeat the comparison by drawing complete lens systems and keeping all accepted images from each drawn system together.  This preserves the observed double, triple, and quadruple systems while using the same mismatch statistic.  Table~I gives the number of complete lens systems needed for the same 95\% separation criterion used in Fig.~\ref{fig:match}(b).
\begin{table}[h]
\centering
\caption{\textbf{Model-separation counts at the system level.}  The first numerical column counts complete lens systems, preserving all accepted images from each system.  The last column gives the corresponding image-waveform count used in Fig.~\ref{fig:match}(b).  The difference between the two columns is the distinction between independent time-separated signals and independent astrophysical lenses.}
\begin{tabular}{lcc}
\toprule
comparison & systems per model & image-level count \\
\midrule
NFW--FDM  & 30 & 60 \\
SIDM--FDM & 50 & 110 \\
NFW--SIDM & -- & -- \\
\bottomrule
\end{tabular}
\end{table}
The system counts are more conservative than the image counts because they retain the shared source and macro-lens parameters.  A full population likelihood would fit all images from one lens at the same time.

\begin{figure*}[t]
    \centering
    \includegraphics[width=\textwidth]{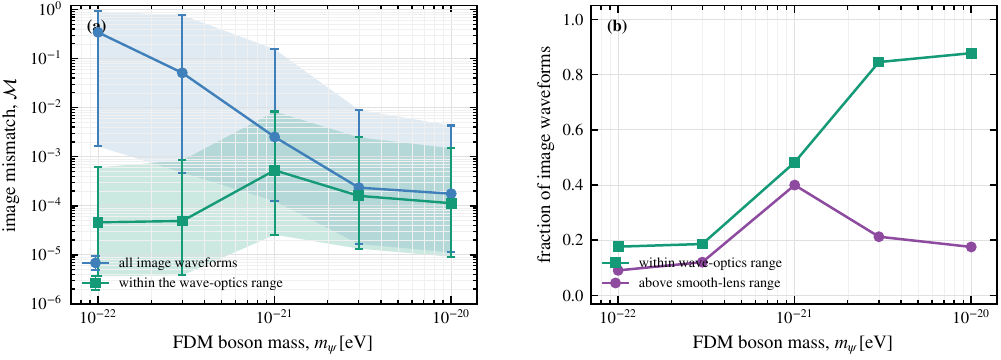}
    \caption{\textbf{Physical scale dependence of the FDM signal.}  The
    same 132 lens systems and 311 time-resolved image waveforms are evaluated
    at five boson masses.  Changing $m_\psi$ changes the density-field
    coherence length while keeping the halo normalization and detector
    calculation fixed.  In (a), blue points show all evaluated image regions
    and green points retain only regions where the local wave-optics
    approximation is valid; bands and error bars give the central 68\%
    range.  In (b), the green squares show the fraction of regions within that
    range and the purple circles show the fraction whose residual is larger
    than the smooth-lens reference.  The result identifies the mass range in
    which FDM leaves the clearest residual; it is not an unconditional
    particle-mass bound.}
    \label{fig:fdmmass}
\end{figure*}
\section{VIII. Dependence on the FDM Boson Mass}
\label{app:massscan}

The main comparison fixes $m_\psi=10^{-21}\,{\rm eV}$.  This choice sets the
spatial coherence scale of the FDM density field through
$\lambda_{\rm dB}\propto m_\psi^{-1}$; it is therefore a physical scale in
the lens, not a free amplitude used to enlarge the waveform effect.  We
test the dependence on this scale by repeating the wave-optics calculation
for
\begin{equation}
m_\psi=10^{-22},\ 3\times10^{-22},\ 10^{-21},\ 3\times10^{-21},\ 10^{-20}\,{\rm eV}.
\end{equation}
The source catalogue, the 132 SIE-plus-shear lens systems, the 311 accepted
image signals, the projected halo population, the FDM field seed, the LISA
orbit, and the first-generation TDI response were held fixed.  Thus the
comparison changes the length scale of the density correlations while
leaving the source population, total projected dark-matter normalization,
and detector calculation unchanged.  No additional FDM amplitude
parameter was introduced.

\begin{figure}[h]
    \centering
    \includegraphics[width=0.48\textwidth]{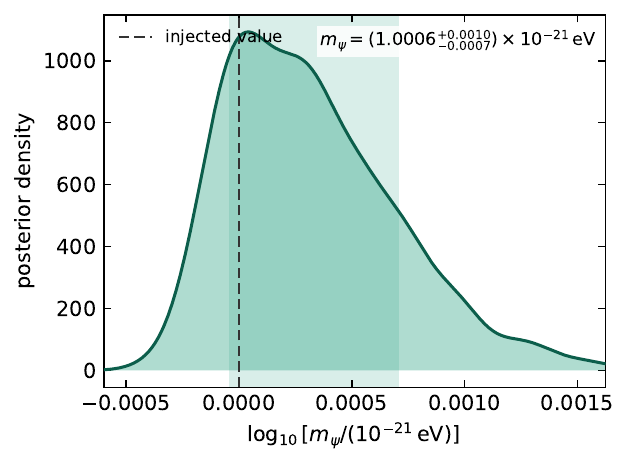}
    \caption{\textbf{A representative FDM-mass measurement.}  The posterior for
    the boson mass from image $6265\_1$, obtained with the same ESA-orbit TDI
    response and smooth-lens waveform used for the population calculation.
    The horizontal coordinate is the boson mass relative to the injected value;
    the shaded region contains 68\% of the posterior probability and the dashed
    line marks the injection.  The narrow concentration shows that the
    frequency-dependent lensing phase carries information about the FDM
    coherence scale.  The result is conditional on the fixed binary, macro-lens,
    and density-field parameters.}
    \label{fig:mpsiposterior}
\end{figure}

Figure~\ref{fig:fdmmass} shows the result.  The blue points in panel (a)
show every evaluated image region, while the green points retain the regions
for which the local wave-optics approximation is valid.  The distinction has a
physical meaning.  At low $m_\psi$, the coherence length is long and the
adopted field normalization produces a large coherent perturbation over many
image regions.  The local approximation then ceases to describe those regions
reliably.
The medians {of the mismatch} for all evaluated regions, $0.34$ at $10^{-22}\,{\rm eV}$ and
$0.051$ at $3\times10^{-22}\,{\rm eV}$, are therefore not interpreted as
measurable distortions: only $17.7\%$ and $18.6\%$ of the image regions
satisfy the validity condition.  We denote this fraction by
\begin{equation}
f_{\rm pass}(m_\psi)\equiv
\frac{N_{\rm pass}(m_\psi)}{N_{\rm image}},
\end{equation}
where $N_{\rm pass}$ is the number of image regions for which the local
wave-optics approximation remains within its range of validity and
$N_{\rm image}=311$ is the number tested at each mass.  Thus $f_{\rm pass}$
tells us how much of the catalogue can be described by this local calculation;
it is not a detection efficiency or a probability that the FDM model is
correct.  At the reference mass, $f_{\rm pass}=48.2\%$ and the mismatch of
the retained images has median
$5.3\times10^{-4}$; the corresponding passing-sample medians
are $1.6\times10^{-4}$ and $1.1\times10^{-4}$ at $3\times10^{-21}$ and
$10^{-20}\,{\rm eV}$, respectively.

Panel (b) gives the physical consequence for model identification: it shows
the fraction of passing image waveforms whose mismatch is above the
smooth-lens reference range.  This fraction is largest near the reference
mass and declines toward higher masses, where the shorter coherence length
causes more of the projected fluctuation to average out in the image-region
integral.  The reference mass gives the strongest residual in this setup;
at higher masses the FDM signal becomes less distinguishable from the
smooth-lens description.  The scan is not a particle-physics posterior,
because it keeps the halo normalization, field realization, and source
catalogue fixed.  Its role is to show why the reference mass was chosen and
which part of the mass range 
{allows for the model-separation.}
A population constraint on $m_\psi$ would require these astrophysical
quantities to be varied and marginalized in the likelihood.

The mass scan tests how the population responds to a change in the FDM
length scale.  We next test whether that scale can be inferred from a
single waveform.  For the representative image $6265\_1$, whose image
signal-to-noise ratio is $1.38\times10^{4}$, we inject
$m_\psi=10^{-21}\,\mathrm{eV}$ and sample
$\log_{10}(m_\psi/\mathrm{eV})$ with a uniform prior from $-22$ to $-20$.
The binary parameters, macro-image solution, FDM density realization, and
detector response are held fixed, so this is a single-event measurement of the
length scale rather than a population constraint.  The likelihood uses the
same frequency-domain waveform, ESA orbit, $A$ and $E$ TDI channels, and
frequency range as in Sec.~V.  The nested sampler uses 512 live points with
a target $\Delta\ln Z=0.05$ and returns 1017 retained posterior samples.  It
gives
\begin{equation}
m_\psi=\left(1.0006^{+0.0010}_{-0.0007}\right)
\times10^{-21}\,\mathrm{eV}
\end{equation}
at 68\% credibility, as shown in Fig.~\ref{fig:mpsiposterior}.  The posterior
is centred on the injected value, showing that the frequency-dependent
FDM phase changes in a measurable way when the coherence length changes.
{This result is complementary to Fig.~\ref{fig:fdmmass}}: the scan shows
where the population is most sensitive, while the posterior demonstrates
how the information appears in an individual high-SNR signal.  Because the
binary, lens, and density-field nuisance parameters are fixed, the quoted
width is conditional.  A physical measurement of the boson mass would have
to marginalize over those quantities and over the lens population.

\section{IX. Limitations}

This calculation isolates one effect: the wave-optics phase that remains after a smooth-lens fit.  A mission analysis should fit all images from one system together and include baryonic perturbers, halos along the line of sight, other macro-lens models, and waveform uncertainties.  Within the model used here, the match retains information about the spatial structure of the perturbing dark matter.

\end{document}